\documentclass{webofc}
\usepackage[varg]{txfonts}   
\woctitle{International Conference on New Frontiers in Physics 2013}
\begin{document}
\title{Applications of dissipative and anisotropic hydrodynamics in description of early stages of relativistic heavy-ion collisions}

\author{Wojciech Florkowski \inst{1,2}\fnsep\thanks{
\email{Wojciech.Florkowski@ifj.edu.pl}} 
}

\institute{Institute of Physics, Jan Kochanowski University, PL-25406~Kielce, Poland
\and
The H. Niewodnicza\'nski Institute of Nuclear Physics, Polish Academy of Sciences,\\ PL-31342 Krak\'ow, Poland
}

\abstract{ Kinetic and hydrodynamic models describing early stages of relativistic heavy-ion
collisions are discussed. We emphasise the role of the shear-bulk coupling for the correct determination of the time
dependence of the bulk viscous pressure.
}
\maketitle
%

\section{Introduction}
\label{intro}

Relativistic hydrodynamics plays an important role  in modeling of relativistic heavy-ion collisions 
\cite{Israel:1979wp,Muronga:2003ta,Baier:2006um,Romatschke:2007mq,Dusling:2007gi,Luzum:2008cw,
Song:2008hj,Denicol:2010tr,Schenke:2011tv, Shen:2011eg,Bozek:2011wa,Niemi:2011ix,Bozek:2012qs}. 
Initially,  approaches based on the perfect fluid hydrodynamics were used. Nowadays, viscous codes
are applied, since i) this allows for better description of  the data and ii) there are general arguments that
the fluid viscosity cannot be zero --- this follows from the quantum mechanical considerations \cite{Danielewicz:1984ww} as well as
from the AdS/CFT correspondence \cite{Kovtun:2004de}.

In this note we analyse simple models describing the early stages of relativistic heavy-ion
collisions and point out difficulties one may encounter in the application of dissipative hydrodynamics~\cite{Martinez:2009mf}.  
We connect these problems with an incomplete character of various computational schemes 
which are used to derive the viscous hydrodynamic equations. Our critical examination of the hydrodynamic 
approaches is based on the comparisons of the hydrodynamic results with the predictions  
of the underlying kinetic theory \cite{Florkowski:2013lza,Florkowski:2013lya,Florkowski:2014sfa}. 

In Refs.~\cite{Florkowski:2013lza,Florkowski:2013lya} we studied the effects connected with the shear viscosity 
and showed that recent formulations of  second-order viscous hydrodynamics  \cite{Denicol:2012cn} agree better with 
the exact solutions of the kinetic equation than the standard Israel-Stewart approach \cite{Israel:1979wp,Muronga:2003ta}. 
In this work we concentrate in more detail on the effects connected with the bulk viscosity \cite{Denicol:2014vaa}.
Recently, it has been found that the finite bulk viscosity coefficient leads to a better description of the flow harmonics 
in ultracentral collisions \cite{Rose:2014fba}. On the theoretical side, it has been shown that the correct description of
the bulk viscous pressure demands the correct treatment of the bulk-viscous coupling 
\cite{Denicol:2014mca,Jaiswal:2014isa}.

The recent methods used to improve the efficacy of the hydrodynamic approaches include: complete second-order 
treatments \cite{Denicol:2012cn}, third-order treatments \cite{El:2009vj,Jaiswal:2013vta}, and anisotropic hydrodynamics  
\cite{Florkowski:2010cf,Martinez:2010sc,Ryblewski:2010bs,Martinez:2010sd,Ryblewski:2011aq,Martinez:2012tu,Ryblewski:2012rr}. 
We argue that the improved description of dissipative processes can be achieved if one uses either the complete second-order 
approaches or anisotropic hydrodynamics (aHydro).

\section{Dissipative and anisotropic hydrodynamics}
\label{dishydro}

A standard hydrodynamic approach is based on the gradient expansion of the underlying phase-space
distribution function around the local equilibrium state described by the Boltzmann (Bose-Einstein or
Fermi-Dirac) distribution function. The corrections to local equilibrium give rise to various 
dissipative currents. At the early stages of heavy-ion collisions, such currents are quite large,
since there exist large gradients present in the system (mainly due to the rapid longitudinal expansion).
Even if the values of the kinetic coefficients are small, the corrections may become large as they are
products of the kinetic coefficients and the gradients. This suggest using a complete second-order
treatment of dissipative hydrodynamics which, in particular, includes the shear-bulk couplings.

The problems of  second-order viscous hydrodynamics \cite{Martinez:2009mf} triggered the development of 
reorganizations of viscous hydrodynamics in which large momentum-space anisotropies are built into 
the leading order of the hydrodynamic expansion \cite{Florkowski:2010cf,Martinez:2010sc,Ryblewski:2010bs,
Martinez:2010sd,Ryblewski:2011aq, Martinez:2012tu,Ryblewski:2012rr}. The newly constructed framework is referred 
to as anisotropic hydrodynamics. {\it It is important to stress that aHydro implicitly  includes transport phenomena and their couplings.}

\section{Kinetic equation for boost-invariant \\ and transversally homogenous systems}
\label{kin}

We have in mind early dynamics of the central rapidity region, hence we assume that the system
is boost-invariant and azimuthally symmetric. In this case, our considerations may be based on 
the simple form of the kinetic equation
\begin{eqnarray}
\frac{\partial f}{\partial \tau}  &=& 
\frac{f^{\rm eq}-f}{\tau_{\rm eq}} \, ,
\label{kineq}
\end{eqnarray} 
where $f(x,p)$ is the phase-space distribution function, $\tau=\sqrt{t^2-z^2}$ is the proper time, and $\tau_{\rm eq}$ is the relaxation time. The requirement of boost invariance implies that $f(x,p)$ may depend only on the three variables: $\tau$, $w$ and $p_T$. The boost-invariant variable $w$ is defined by the expression $w =  tp_L - z E$, where $p_L$ ($p_T$) is the longitudinal (transverse) momentum and $E$ is the energy. The equilibrium background distribution function  $f^{\rm eq}$ may be written as
\begin{eqnarray}
f^{\rm eq}(\tau,w,p_T) =
\frac{2}{(2\pi)^3} \exp\left[
- \frac{\sqrt{w^2+p_T^2 \tau^2}}{T(\tau) \tau}  \right],
\end{eqnarray}
where $T$ is an effective temperature.

The first moment of the kinetic equation defines the divergence of the energy-momentum tensor that should be conserved. 
\begin{eqnarray}
T^{\mu\nu}(\tau) = g_0 \int dP \, p^\mu p^\nu f(\tau,w,p_T), \qquad \partial_\mu T^{\mu\nu} = 0. 
\label{Tmunu1}
\end{eqnarray}
Using the symmetry properties of the distribution function, we rewrite (\ref{Tmunu1}) in the form 
\cite{Florkowski:2010cf,Martinez:2012tu}
\begin{eqnarray}
T^{\mu\nu} = ({\cal E} + {\cal P}_T) u^\mu u^\nu - {\cal P}_T g^{\mu\nu} + ({\cal P}_L-{\cal P}_T) z^\mu z^\nu,
\label{Tmunu2}
\end{eqnarray}
where $u^\mu=(t,0,0,z)/\tau$ and $z^\mu=(z,0,0,t)/\tau$. The energy density and the two (longitudinal and transverse) pressures are defined as the integrals over the distribution function multiplied by the appropriate combinations of the momentum.  The parameter $g_0$  in Eq.~(\ref{Tmunu1}) is the degeneracy factor connected with internal degrees of freedom different than spin.

The initial condition at the time $\tau=\tau_0$ used to solve numerically Eq. (\ref{kineq}) 
corresponds to the Romatschke-Strickland (RS) form of the distribution function \cite{Romatschke:2003ms} 
\begin{eqnarray}
f_0(w,p_T) &=&  \frac{1}{4\pi^3}
\exp\left[
-\frac{\sqrt{(1+\xi_0) w^2 + p_T^2 \tau_0^2}}{\Lambda_0 \tau_0}\, \right].
\nonumber \\
\label{RS}
\end{eqnarray}
This form reduces to an isotropic Boltzmann distribution if the anisotropy parameter $\xi_0=\xi(\tau_0)$ vanishes. In this case, the transverse momentum scale $\Lambda_0=\Lambda(\tau_0)$ is equal to the system's initial temperature $T_0$. The methods of solving Eq.~(\ref{kineq}) has been expained in more detail in \cite{Florkowski:2013lza,Florkowski:2013lya,Florkowski:2014sfa}. 

\begin{figure}
\centering
\includegraphics[width=1cm,width=0.495\textwidth]{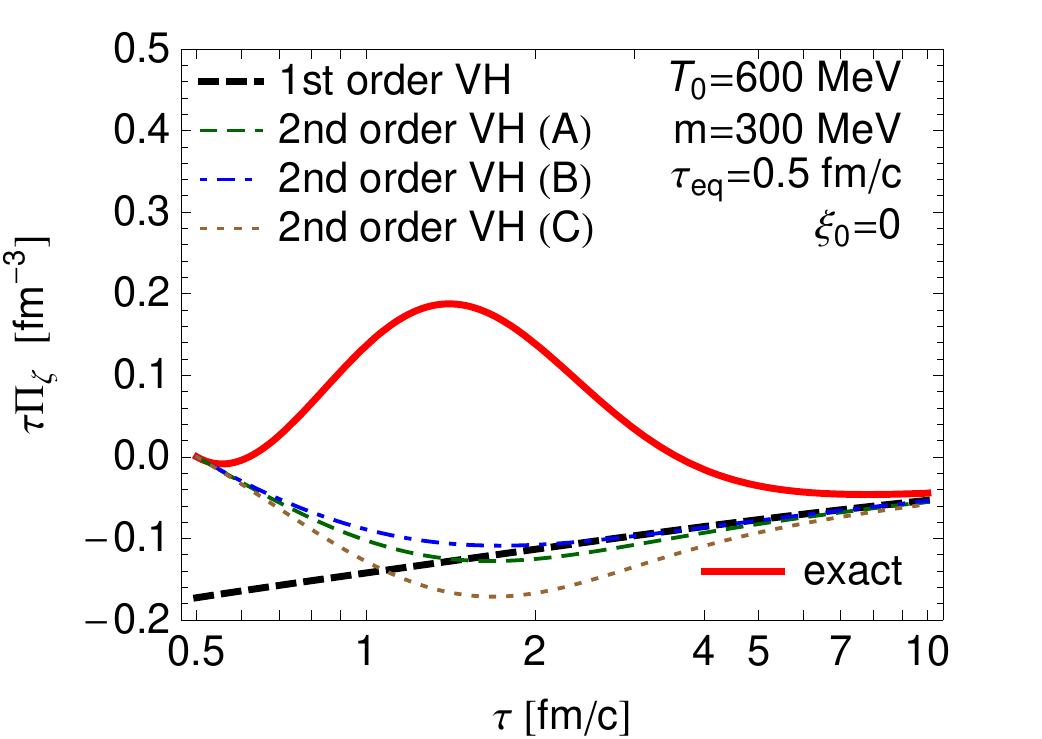} 
\includegraphics[width=1cm,width=0.495\textwidth]{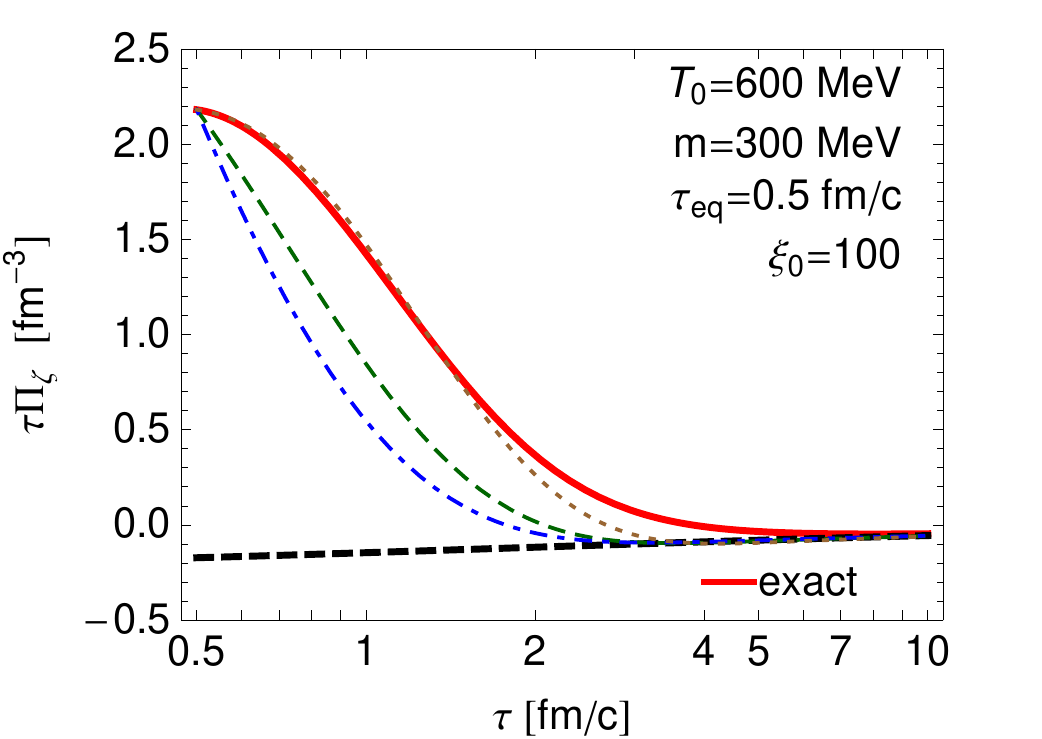}
\caption{Time dependence of the bulk viscous pressure (multiplied by the proper time) obtained from the kinetic theory and various hydrodynamic equations.}
\label{fig-1}      
\end{figure}

If the solution of the kinetic equation is found, one can calculate the bulk viscous pressure from the equation
\begin{eqnarray}
 \Pi^{k}_{\zeta}(\tau) = \frac{1}{3}\left[{\cal P}_{\|}(\tau)+2 {\cal P}_{\bot}(\tau)-3{\cal P}_{\rm eq}(\tau)  \right],
\end{eqnarray}
where the equilibrium pressure ${\cal P}_{\rm eq}$ is connected with the energy density ${\cal E}$ by the equation of state (for the massive gas). In Fig.~\ref{fig-1} the red solid curves show the time dependence of the bulk viscous pressure calculated from the kinetic equation for the two different initial conditions characterised by the momentum anisotropy parameter $\xi_0$. In the two cases the initial temperature of the system is $T_0$~=~600~MeV, the effective particle mass is $m$~=~300~MeV, and the equilibration time $\tau_{\rm eq}$=~0.5~fm.

\section{Bulk viscous pressure in dissipative and anisotropic hydrodynamics}
\label{sect:Pi}

In the most popular formulations of dissipative hydrodynamics, the bulk viscous pressure evolution
is determined by one of the following three equations:
\begin{eqnarray}
\tau_\Pi \dot{\Pi}_\zeta + \Pi_\zeta &=& 
-\frac{\zeta}{\tau} 
-\frac{1}{2} \tau_\Pi \Pi_\zeta \left[
\frac{1}{\tau} -\left( \frac{\dot{\zeta}}{\zeta} +\frac{\dot{T}}{T} \right)
\right], \label{A}
\end{eqnarray}
\begin{eqnarray}
\tau_\Pi \dot{\Pi}_\zeta + \Pi_\zeta &=& 
-\frac{\zeta}{\tau} - \frac{4}{3} \tau_\Pi \Pi_\zeta \frac{1}{\tau} ,
\label{B}
\end{eqnarray}
\begin{eqnarray}
\tau_\Pi \dot{\Pi}_\zeta + \Pi_\zeta &=& 
-\frac{\zeta}{\tau} .
\label{C}
\end{eqnarray}
We have solved Eqs.~(\ref{A})--(\ref{C}) with the initial conditions matching those used in 
the kinetic-theory approach. The results are shown in Fig.~\ref{fig-1}. The cases (A), (B), 
and (C) correspond to Eqs.~(\ref{A}), (\ref{B}), and (\ref{C}), respectively. The thick
dashed line describes the first order hydrodynamic result where the bulk pressure
is directly connected with the bulk viscosity through the formula $\Pi_\zeta = -\zeta/\tau$.
{\it The striking result  of our comparisons is that none of the hydrodynamic equations (\ref{A})--(\ref{C}) 
can properly reproduce the numerical result.}

\begin{figure}
\centering
\includegraphics[width=1cm,width=0.6\textwidth]{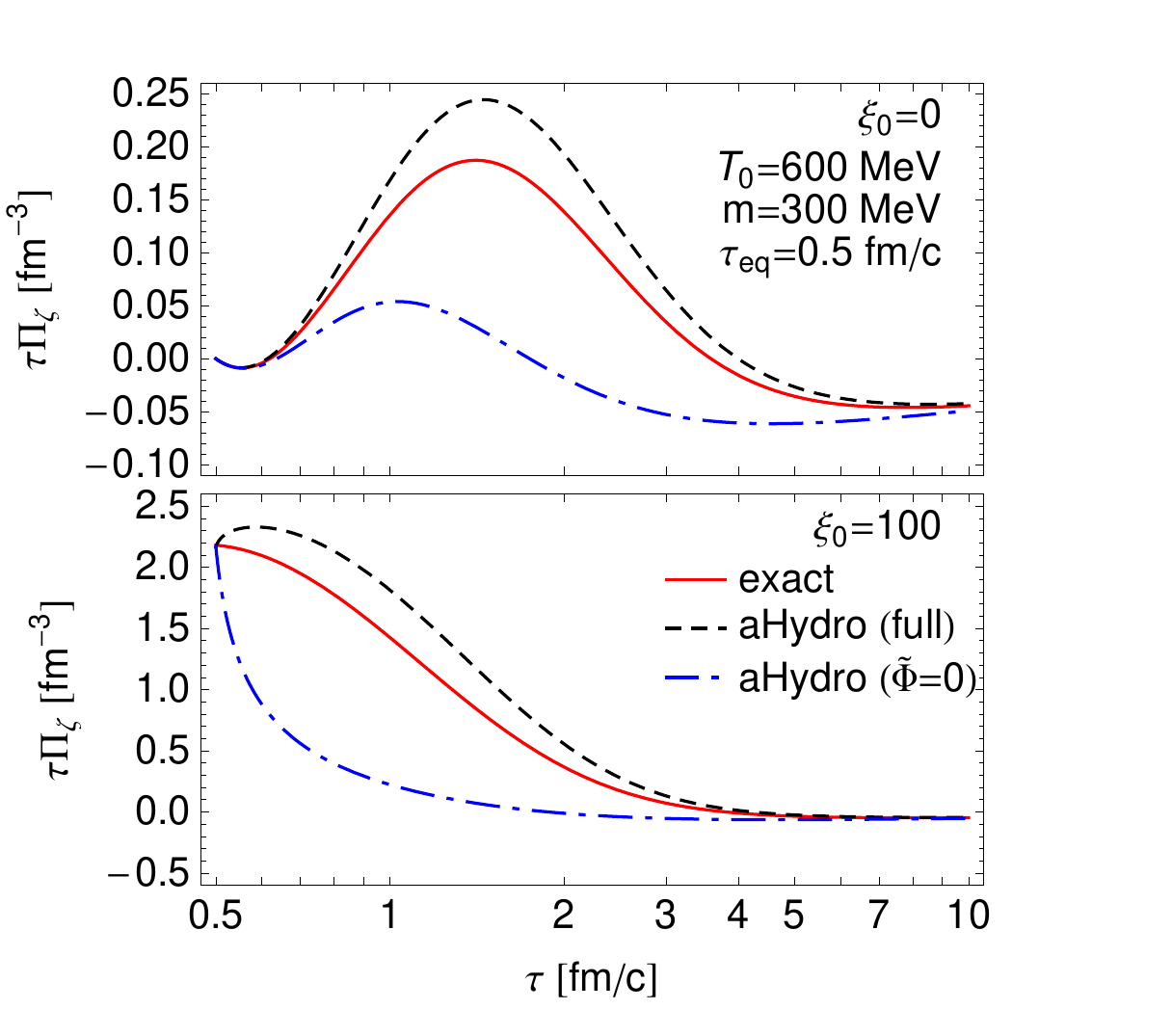} 
\caption{Time dependence of the bulk viscous pressure for two different initial anisotropy  parameters: $\xi_0=0$ (upper panel) and $\xi_0=100$ (lower panel). The solid lines show the exact result. The dashed-dotted lines represent the predictions of aHydro 
formulated in Ref.~\cite{Florkowski:2014bba}, while the dashed lines represent the predictions of Ref.~\cite{Nopoush:2014pfa}.}
\label{fig-2}      
\end{figure}

In the aHydro approach defined in Ref.~\cite{Florkowski:2011jg,Florkowski:2014bba}, 
one assumes that the leading-order phase-space 
distribution function has always the RS form with the parameters $\xi$ and $\Lambda$ depending 
on the proper time~$\tau$.  The dynamic equations are constructed from the first and second moments of 
the Boltzmann equation~(\ref{kineq}).  In a more recent formulation of aHydro \cite{Nopoush:2014pfa}, the
leading-order phase-space distribution function depends additionally on a parameter ${\tilde \Phi}$,
and the dynamic equations follow from the zeroth, first and second moments of the kinetic 
equation~(\ref{kineq})~\footnote{There exist also other formulations of aHydro, where some part of
anisotropy is treated perturbatively by adding corrections to the leading-order term, see Ref.~\cite{Bazow:2013ifa}}.
The results obtained with the two versions of aHydro are shown in
Fig.~\ref{fig-2}, where we compare them with the results of the kinetic theory.  One observes that the
inclusion of an additional parameter in the ansatz for the distribution function helps to get
a reasonable agreement with the kinetic theory. We also show that aHydro describes the evolution
of the bulk pressure much better than the dissipative hydrodynamics based on 
Eqs.~(\ref{A}), (\ref{B}), or (\ref{C}).

\begin{figure}
\centering
\includegraphics[width=1cm,width=0.495\textwidth]{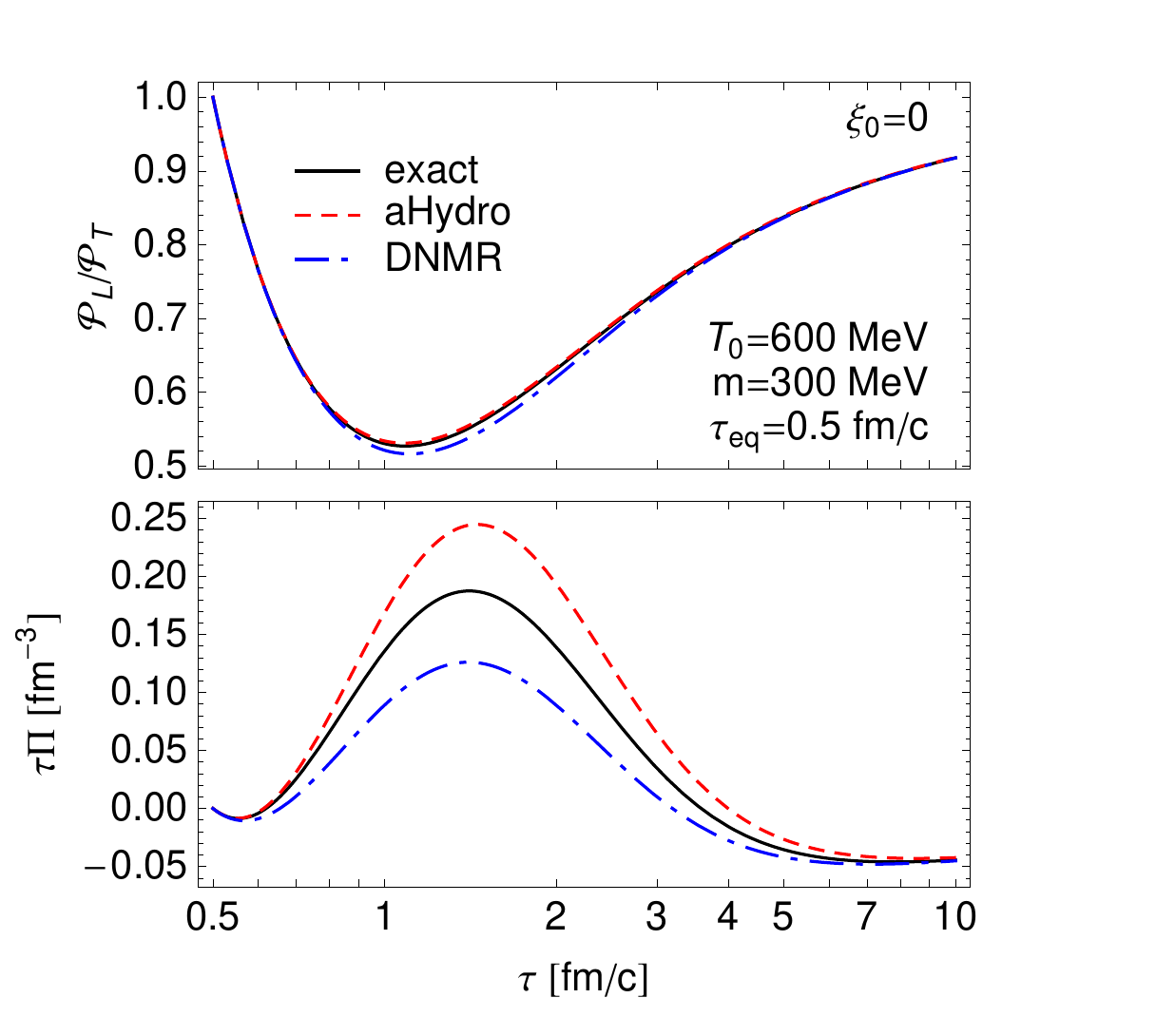} 
\includegraphics[width=1cm,width=0.495\textwidth]{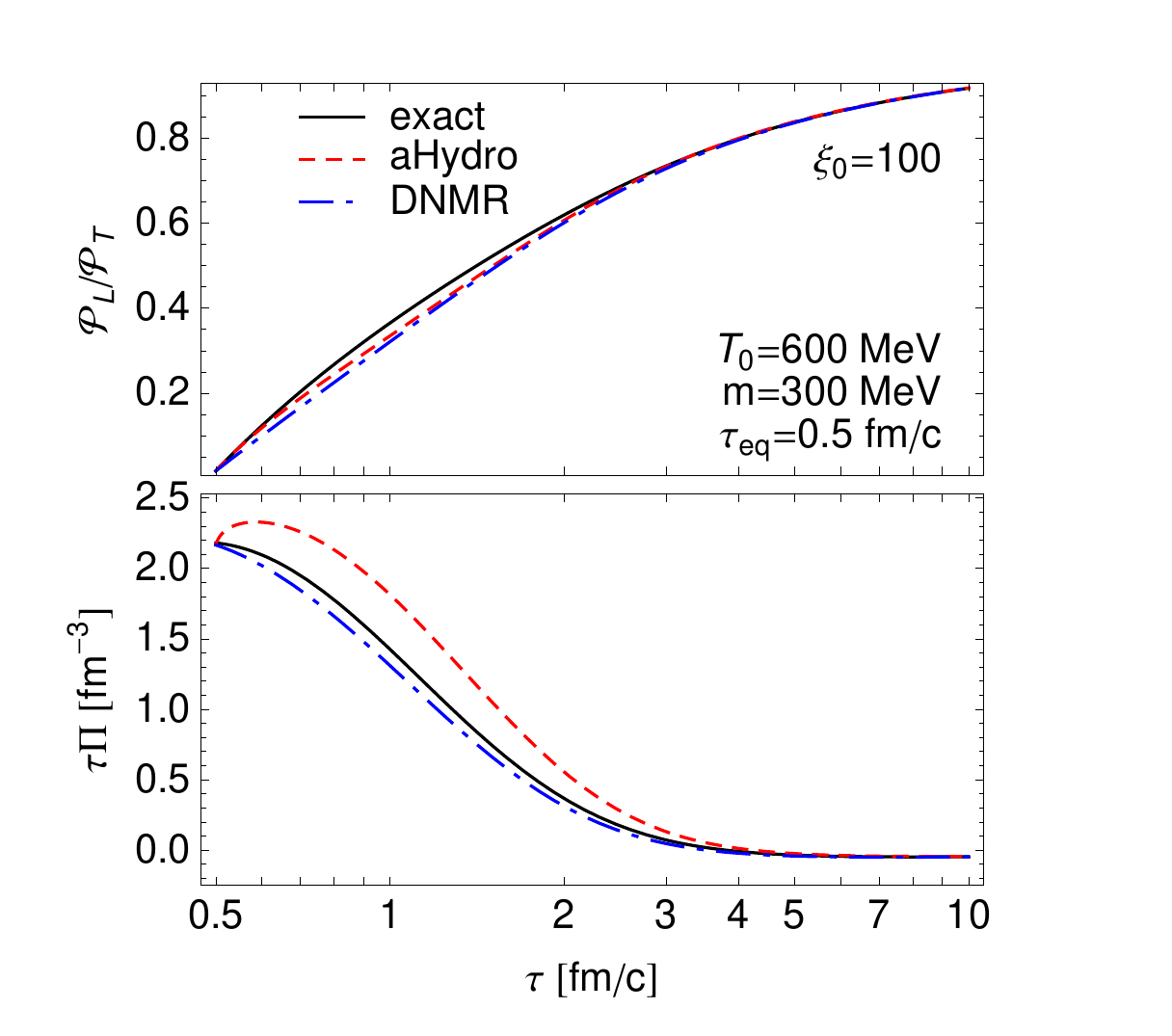}
\caption{Time dependence of the ratio of the longitudinal and transverse pressure (upper panels) and of the bulk viscous pressure (lower panels). The left panels describe the cases where the initial anisotropy parameter vanishes, the right panels describe the cases where $\xi_0=100$.  The kinetic theory results are shown as the solid lines, the complete second order hydrodynamic calculations including the shear-bulk coupling are represented by the dashed-dotted lines, and the aHydro results are shown as the dashed lines. }
\label{fig-3}      
\end{figure}

Discrepancies between the results of the kinetic theory and dissipative hydrodynamics may
be connected with the absence of the shear-bulk coupling in Eqs.~(\ref{A}), (\ref{B}), and (\ref{C}).
In Fig.~\ref{fig-3} we show the results for the bulk pressure with the shear-bulk coupling included
\cite{Denicol:2014mca} (dashed-dotted lines) and make comparisons with the kinetic theory results 
(solid lines ) and aHydro results \cite{Nopoush:2014pfa} (dashed lines). The inclusion of the shear-bulk 
coupling improves the agreement between dissipative hydrodynamics and the exact kinetic theory solution.
Similar conclusions have been also reached in Ref.~\cite{Jaiswal:2014isa}.

\section{Conclusions}
\label{sect:res}

Detailed comparisons between the exact results of the kinetic theory and the predictions of hydrodynamic
models allow us to select the right structure of the hydrodynamic equations and the correct form
of the kinetic coefficients. In this way we select the appropriate structure of the hydrodynamic equations that
may be used to model relativistic heavy-ion collisions. We emphasise that the shear-bulk coupling is crucial 
for the correct determination of the time dependence of the bulk viscous pressure.

In this note we have presented the analysis of one-dimensional systems. Very recently, the exact solutions 
of the two-dimensional systems have become also available \cite{Denicol:2014xca,Denicol:2014tha} 
(for the systems which are boost-invariant and azimuthally symmetric) but they are restricted to conformal systems.  
Therefore, they are suitable for studies of the effects connected with shear viscosity.

\bigskip
{\bf Acknowledgments:}  I thank Gabriel Denicol, Ewa Maksymiuk, Radoslaw Ryblewski, Michael Strickland, and Leonardo Tinti for fruitful collaborations which have led us to the results presented in this paper.  This work was supported in part by the Polish National Science Center with decision No.~DEC-2012/06/A/ST2/00390.


\end{document}